\documentclass[reprint, amsmath, amssymb, aps, prl, superscriptaddress, floatfix, nofootinbib, citeautoscript, longbibliography]{revtex4-1}
\usepackage[utf8]{inputenc}
\usepackage[T1]{fontenc}
\usepackage[parfill]{parskip}
\usepackage{siunitx}
\usepackage{blindtext}
\usepackage[colorlinks, linkcolor=blue, citecolor=blue, urlcolor=blue, breaklinks=true]{hyperref}
\usepackage{graphicx}
\usepackage{dcolumn}
\usepackage{bm}
\usepackage{color}
\usepackage{units}
\usepackage{mathtools}
\usepackage{dsfont}
\usepackage{hyperref}
\usepackage{xcolor}
\usepackage{float}

\newcommand{\expec}[1]{\left\langle #1 \right\rangle}

\newcommand{\ket}[1]{\left| #1 \right\rangle}

\begin{document}

\title{Photonic angular super-resolution using twisted N00N states}

\author{Markus Hiekkam\"aki}
\email{markus.hiekkamaki@tuni.fi}
\affiliation{Tampere University, Photonics Laboratory, Physics Unit, Tampere, FI-33720, Finland}

\author{Fr\'ed\'eric Bouchard}
\affiliation{National Research Council of Canada, 100 Sussex Drive, Ottawa, Ontario K1A 0R6, Canada}

\author{Robert Fickler}
\email{robert.fickler@tuni.fi}
\affiliation{Tampere University, Photonics Laboratory, Physics Unit, Tampere, FI-33720, Finland}

\begin{abstract}
The increased phase sensitivity of N00N states has been used in many experiments, often involving photon paths or polarization. 
Here we experimentally combine the phase sensitivity of N00N states with the orbital angular momentum (OAM) of photons up to 100\,$\hslash$, to resolve rotations of a light field around its optical axis. 
The results show that both a higher photon number and larger OAM increase the resolution and achievable sensitivity.
The presented method opens a viable path to unconditional angular super-sensitivity and accessible generation of N00N states between any transverse light fields.
\end{abstract}

\maketitle

During the past few decades, N00N states have been the focus of several studies where their potential has been explored in different metrological applications \cite{rarity1990two, kuzmich1998sub,dowling2008quantum, polino2020photonic}.
Specifically, a N00N state refers to an extremal superposition of $N$ quanta between two orthogonal modes, i.e., $\frac{1}{\sqrt{2}}(\ket{N,0} + \ket{0,N})$
\cite{dowling2008quantum}.
These states owe their usefulness to an increased phase sensitivity that an $N$-photon Fock state has in comparison to a single photon, or more classical states of light.
The increased phase sensitivity means that a phase $\phi$ affects the Fock state $\ket{N}$ $N$ times, changing the state to $e^{\textit{i}N\phi}\ket{N}$, whereas a coherent state $\ket{\alpha}$ would only gain the phase $e^{\textit{i}\phi}\ket{\alpha}$ \cite{dowling2008quantum}.
This increase in phase sensitivity has been utilized in many proof-of-principle experiments, most commonly by preparing two photons in a superposition of two paths \cite{rarity1990two} or polarizations \cite{kuzmich1998sub}.
One notable example is the demonstration of an unconditional quantum advantage in sensitivity, using two-photon polarization N00N states \cite{slussarenko2017unconditional}.

Similarly to the phase sensitivity scaling with photon number, the sensitivity in rotation measurements around the optical axis scales with the amount of helical twistedness in the wavefront of the light used \cite{barnett2006resolution}.
This sensitivity is related to the rotational symmetry of the helically twisted wavefront of a light beam with non-zero orbital angular momentum (OAM).
The amount of wavefront twistedness, or OAM, a photon can have is quantized to integer multiples $\ell$ of $\hslash$, and is theoretically unbounded \cite{erhard2018twisted}, leading to a theoretically unbounded increase in measurement sensitivity.
Experimentally, values of up to 10010 quanta of OAM were already demonstrated \cite{fickler2016quantum}, however, this value is bounded by the finite aperture of the optical system \cite{padgett2015divergence}. 

Theoretical and experimental studies have examined methods of combining the increased phase sensitivity of quantum states and the optimal rotation sensitivity of light beams with large OAM \cite{jha2011supersensitive,fickler2012quantum,liu2018squeezing, d2013photonic, ming2014generation}.
In these studies however, instead of experimentally implemented twisted N00N states, the authors used either squeezed light states, light directly from a spontaneous parametric down conversion (SPDC) source, or multiple paths for the photons with different OAM values to travel.
These implementations lack the robustness and simplicity of a single-path operation which can be achieved with the recently introduced method for bunching photons into different OAM N00N states \cite{hiekkamakiHighDim}.

In this study, we experimentally demonstrate the increased rotation sensitivity of twisted N00N states by adapting the method demonstrated in \cite{hiekkamakiHighDim}.
With this method, we are able to show the increased rotation sensitivity using N00N states with photon numbers 1 and 2, and OAM values up to $100\,\hslash$.
Our results show that twisted two-photon N00N states have the potential for an angular uncertainty scaling $\propto \nicefrac{1}{\ell \times N}$, whereas classical light is limited to a scaling $\propto \nicefrac{1}{\ell \times \sqrt{N}}$ \cite{jha2011supersensitive,d2013photonic}. 
Although the amount of OAM is limited by the physical aperture, increasing the number of photons in a twisted N00N state has the potential to surpass any classical angular resolution limit.
Due to the simplicity of the presented method, spatial mode N00N states with large OAM values and high efficiencies are achievable even with current technologies.
As such, our work opens up novel ways to generate N00N states invoking the transverse spatial degree of freedom and offers a path to unconditional angular super-sensitivty.

\begin{figure*}[ht]
    \centering
    \includegraphics[width = \textwidth]{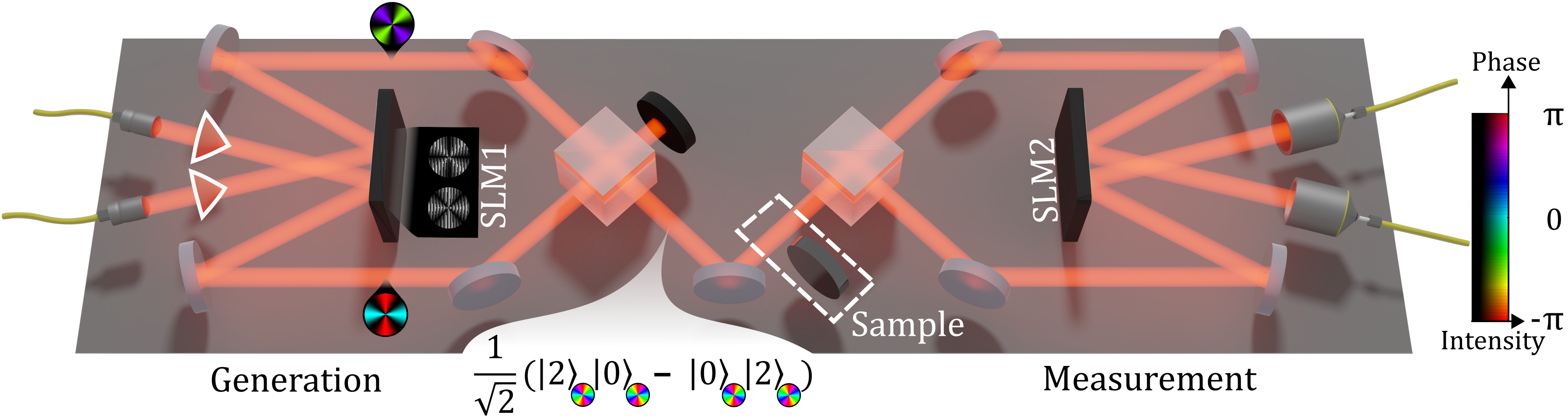}
    \caption{Conceptual image of the experimental setup. Two holograms are used on the first spatial light modulator (SLM1) to imprint the wanted structures onto each photon, independently.
    The two photons are then overlapped using a beamsplitter, to enable photon bunching into the same spatial structure; hence, allowing for a single beam operation when probing a sample.
    To measure the two-photon state, the photons are separated with a beamsplitter and sent to SLM2 where another set of holograms are used to measure the structure of each photon (see main text for details).
    The insets show an example of the holograms displayed on SLM1 to generate an OAM N00N state with $\ell = \pm 2$ and an example of a sample position.
    Additionally, the insets show the generated spatial amplitude and phase structures the photons have at different points of the setup, visualized by a two-dimensional color map (see color bar on right).}
    \label{fig:setup}
\end{figure*}

To create a two-photon twisted N00N state, two photons that have orthogonal transverse-spatial structures, but are otherwise indistinguishable, need to be brought into the same beam path.
A unitary that transforms the modes into a mutually unbiased basis (MUB), i.e., a Hadamard operation $\hat{H}_2$, then leads to a bunching of the two photons into the original spatial structures \cite{hiekkamakiHighDim}. This interference is analogous to the well-known Hong-Ou-Mandel interference realized by a beamsplitter transformation \cite{HOM_orig}.
However, since the beamsplitter-like transformation $\hat{H}_2$ is unitary, the two photons stay orthogonal in a certain basis
\begin{equation*}
    \hat{H}_2\ket{1,1}_{\ell,-\ell} = \frac{1}{\sqrt{2}}\left(\ket{2,0}_{\ell,-\ell} - \ket{0,2}_{\ell,-\ell}\right) = \ket{1,1}_{M_1,M_2},
\end{equation*}
where $M_i$ refer to the modes of another MUB of the OAM modes $\{\ell,-\ell\}$.
Because of this feature, it is possible to include the beamsplitter-like operation into the state generation, before bringing them into the same beam path, while still achieving the same two-photon bunching.
Hence, the same two-photon twisted N00N state is generated.

To experimentally verify the efficacy of this method, we use a spontaneous parametric down conversion (SPDC) source to generate photon pairs and the setup shown in Fig.~\ref{fig:setup} (see the Supplementary for more details).
The photon pair is coupled out of single-mode fibers (SMFs) onto two separate regions of a spatial light modulator (SLM), as shown on the left side of Fig.~\ref{fig:setup}, where the photons are structured using holographic phase and amplitude modulation \cite{bolduc2013exact,forbes2016creation}.
The structured photons are then overlapped with a beamsplitter to bring them into the same path and enable bunching into OAM structures.
To measure the two-photon state, a second beamsplitter probabilistically separates the photons, and a second SLM (SLM2) is used in conjunction with two SMFs to filter the spatial structures of the photons independently \cite{bouchard2018measuring, mair2001entanglement}.
Both of the SLMs that were used, were wavefront corrected using the method specified in Ref.~\cite{jesacher2007wavefront}.
For single-photon N00N states, only one input and output fiber were used and the other photon was detected at the two-photon source, to herald a single-photon state \cite{hong1986experimental}.

To confirm that the photons bunch into a N00N state, we first prepare a two-photon N00N state with an OAM value of $\ell = \pm1$, and verify its quantum correlations using an entanglement witness \cite{guhne2009entanglement,fickler2014quantum}. 
Measuring the state in all three MUBs, we achieve a witness value of $w = 2.92 \pm 0.02$, which is greater than the maximal value of $w = 1$ for separable states and close to the maximum value of $w = 3$ of the witness for maximally entangled states. 
 
After this initial confirmation, we proceed to examine the angular resolution and sensitivity of these OAM N00N states using our measurement scheme.
In these measurements, we prepare heralded single photons and two-photon N00N states with OAM values of $|\ell| = \{1, 2, 3, 5, 10, 25, 50, 100\}$.
For OAM values $|\ell| < 10$, we use mode carving \cite{bolduc2013exact} and intensity flattening \cite{bouchard2018measuring} to create and measure the structures, respectively.
The amplitude modulation implemented in these procedures is needed to get as close as possible to the MUB states of OAM light fields. 
The OAM states have a complex field structure $E_\ell(\theta) \propto e^{\text{i} \ell \theta}$,
where $\theta$ is the azimuthal coordinate.
Hence, the MUB structures are of the form $E_{M_1/M_2}(\theta) \propto \left(e^{\text{i} \ell \theta} \pm e^{-\text{i} \ell \theta} \right)$.
These MUB states are often called petal beams \cite{litvin2014doughnut}.
Examples of these structures for $\ell = \pm 2$ are shown in the insets of Fig.~\ref{fig:setup}.
When generating photons with OAM $|\ell| \geq 10$, no amplitude modulation is required as the spatial structures are sufficiently filtered by the limited aperture of our system.
Thus, simple phase imprinting \cite{forbes2016creation} and phase flattening \cite{mair2001entanglement} are used to generate and measure the desired modes, respectively.

\begin{figure}[htb]
    \centering
    \includegraphics[width = 0.5\textwidth]{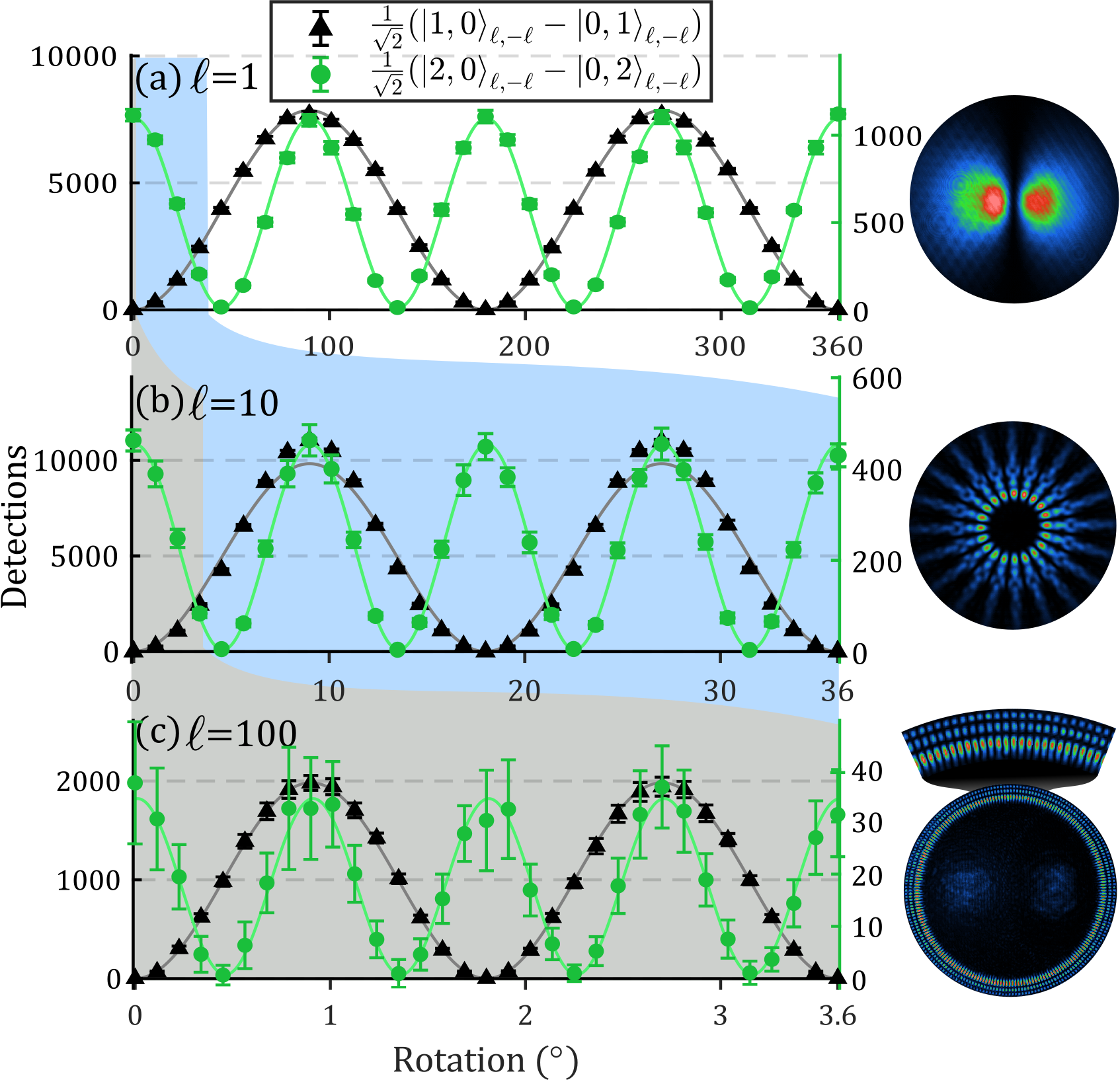}
    \caption{Detected single photons and two-photon coincidences as a function of rotation angle. 
    The single photons were prepared in the modes shown in the insets (insets show false-color images of structures taken with camera and laser light) and the corresponding two-photon N00N states were created by imprinting the same structure on one photon and its orthogonal pair (same structure rotated by $\nicefrac{180^\circ}{2\ell}$) on the other. (a), (b), and (c) show the measurements for single (two-photon) counts with integration times of 2\,s (3\,s), 1\,s (3\,s), and  2\,s (8\,s) and OAM values of $\ell = 1,10,$ and $100$, respectively.
    The error bars have been calculated as standard deviations from at least 19 repetitions of the measurement at each point, and the solid lines are fits of the form shown in eq.~\eqref{eq:fit}.
    The decreased period between oscillations shows the angular super-resolution achieved with the two-photon N00N states \cite{resch2007time}. 
    For the two-photon measurements, accidental coincidences have been subtracted.
    }
    \label{fig:SineData}
\end{figure}

To demonstrate angular super-resolution, we simulate the rotation of our photon structures by rotating the measurement holograms on SLM2.
As with the more common two-photon Mach-Zehnder interferometer, a second Hadamard transformation is needed to detect the phase change.
For the path degree of freedom, the second Hadamard transformations correspond to another beamsplitter; in the case of OAM modes, the second transformation can be performed in the measurement by simply projecting the photons onto the petal mode basis (see Supplementary for details).
Hence, the two-photon state was measured by projecting the photons on orthogonal petal structures, which can only result in an interference curve with perfect visibility in the case of bunching.
Interestingly, projecting both photons on identical petal structures can produce a perfect fringe visibility irrespective of bunching, however, with an increased amplitude in the case of bunching (see Supplementary). 

Since rotating the light field by an angle $\varphi$ induces a phase $e^{\text{i} \ell N \varphi}$ on an N-photon Fock state \cite{jha2011supersensitive}, the theoretically expected detection rate is 
$\expec{\hat{M}} = \frac{M}{2}\left( 1 - \cos \left( 2N\ell \varphi \right) \right)$,
where $M$ is the number of repetitions of the measurement and $N$ is the number of photons used in the N00N state.
From the detection rate, the theoretical scaling of the angular uncertainty can be expressed as 
\begin{equation}
    \label{eq:precision_scaling}
   \left|\Delta \varphi \right| = \frac{\expec{\Delta {\hat{M}}}}{\left| \nicefrac{\partial \expec{\hat{M}}}{\partial \varphi}\right|} = \frac{1}{2\sqrt{M}N\ell}
\end{equation}
For a detailed derivation based on Refs. \cite{jha2011supersensitive,scully1993quantum}, see the Supplementary.

The rotation measurements with $\ell = \{\pm1, \pm10, \pm100\}$ are shown in Fig.~\ref{fig:SineData}. 
The figure shows that increasing the amount of OAM increases the achievable resolution, and changing from a single-photon to a two-photon N00N state doubles the resolution.

To further analyze the measured data, we estimate the Fisher information and angular precision for each measurement.
Therefore, we first fit a curve to each set of the measured data using a weighted nonlinear least squares fit (each point is weighted by the reciprocal of the measured variance).
The fitted curve is 
\begin{equation}
    \label{eq:fit}
    \frac{A}{2}\left(1-\cos \left(2N\ell\frac{\pi}{180^\circ} \varphi-c \right)\right) + D,
\end{equation}
 where $A$ is the amplitude of the cosine curve, $D$ is the offset, and $c$ sets the position of $0^\circ$ rotation.
Hence, $\frac{A}{A+2D}$ gives an estimate of the visibility of the curve, based on the fit.
For the single-photon measurements, we obtain an average visibility of 0.999, whereas for the two-photon measurements the corresponding value is 0.956 averaged over all measurements.
The maximum standard error for the visibilities is 0.011, calculated for the two-photon $\ell = \pm 100$ state, from the confidence intervals of the fitting parameters.

From these fits, we are able to estimate the expected Fisher information $F$ (see supplementary for more information),
and angular uncertainty
\begin{equation}
    \label{eq:precision}
    \left|\Delta \varphi \right| = \frac{\Delta M(\varphi)}{A N\ell \frac{\pi}{180^\circ} \left| \sin\left( 2 N\ell \frac{\pi}{180^\circ} \varphi -c \right) \right|},
\end{equation}
where $\Delta M(\varphi)$ is the standard deviation for each measurement angle calculated from around 25 repetitions, depending on the photon number and OAM value. 
The Fisher information and angular precision, calculated from the fits, are shown in Fig.~\ref{fig:Fisher} for $\ell = 100$ and in the supplementary material for $\ell = 1$.

\begin{figure*}[htb]
    \centering
    \includegraphics[width = 1\textwidth]{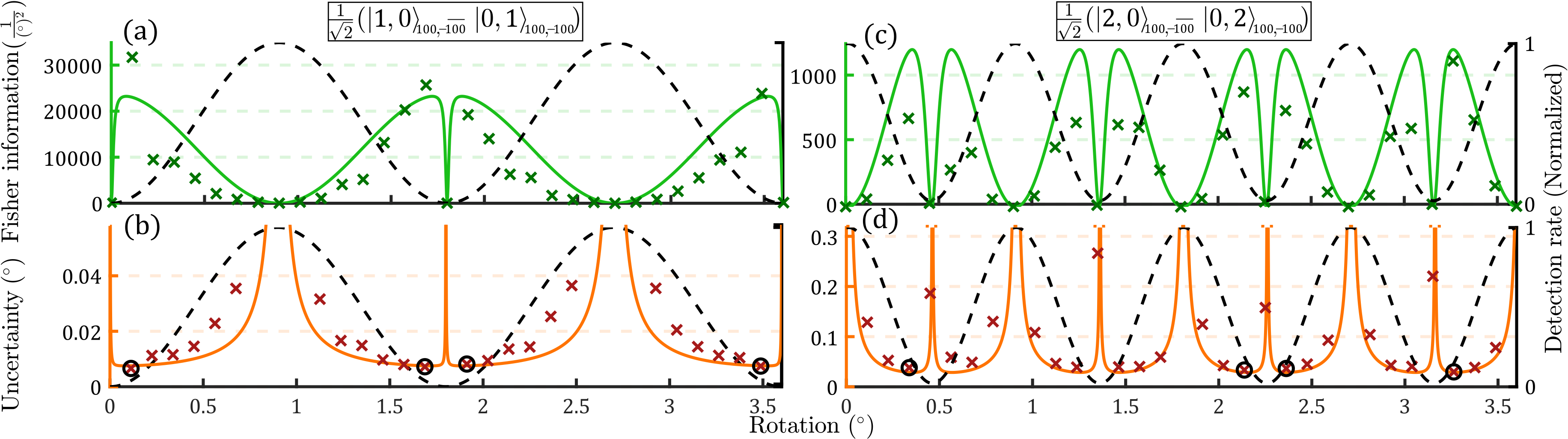}
    \caption{Fisher information and angular uncertainty for $\ell = 100$ N00N states.
    On the upper row, the continuous green line is the Fisher information $F$ multiplied by the estimate for the total number of heralded single photons (or two-photons) before losses.
    The green crosses are the reciprocal of the variance calculated from eq.~\eqref{eq:precision}.
    On the bottom row, the continuous curves are calculated using eq.~\eqref{eq:precision} and Poissonian errors calculated from the fit. 
    The red crosses are the experimentally determined uncertainties, calculated using eq.~\eqref{eq:precision}.
    Plots (a) and (b) display the heralded single photon data, and (c) and (d) contain two-photon data.
    In all graphs, the black dashed lines depict the interference curves for reference.
    On the bottom row, the uncertainty values that have been circled are used for calculating the respective sensitivities in Fig.~\ref{fig:Sensitivities}.}
    \label{fig:Fisher}
\end{figure*}

Plots (a) and (c) in Fig.~\ref{fig:Fisher} show that the expected Fisher information curves follow the reciprocal of the rotation angle variance, meaning that the results are close to the Cram\'er-Rao bound of our specific state measurement \cite{polino2020photonic}. 
Similarly, the expected angular uncertainties mostly agree with the measured angular uncertainties.
This indicates that the achieved precision is close to the maximum precision bounded by Poissonian noise.
The differences between the expected curves and measured data arise from a few sources, namely the limited number of repetitions used to calculate the standard deviation, the decoupling of the system during long measurements, and differences in system losses due to different bandwidths of our single-photon source and the laser used in characterization.
The change in precision caused by errors that were larger than Poissonian, is especially apparent with the $\ell = 100$ measurements where a small drift over time has a comparatively large effect on the alignment of the small structures of the spatial structure.

\begin{figure}[hb]
    \centering
    \includegraphics[width = 0.45\textwidth]{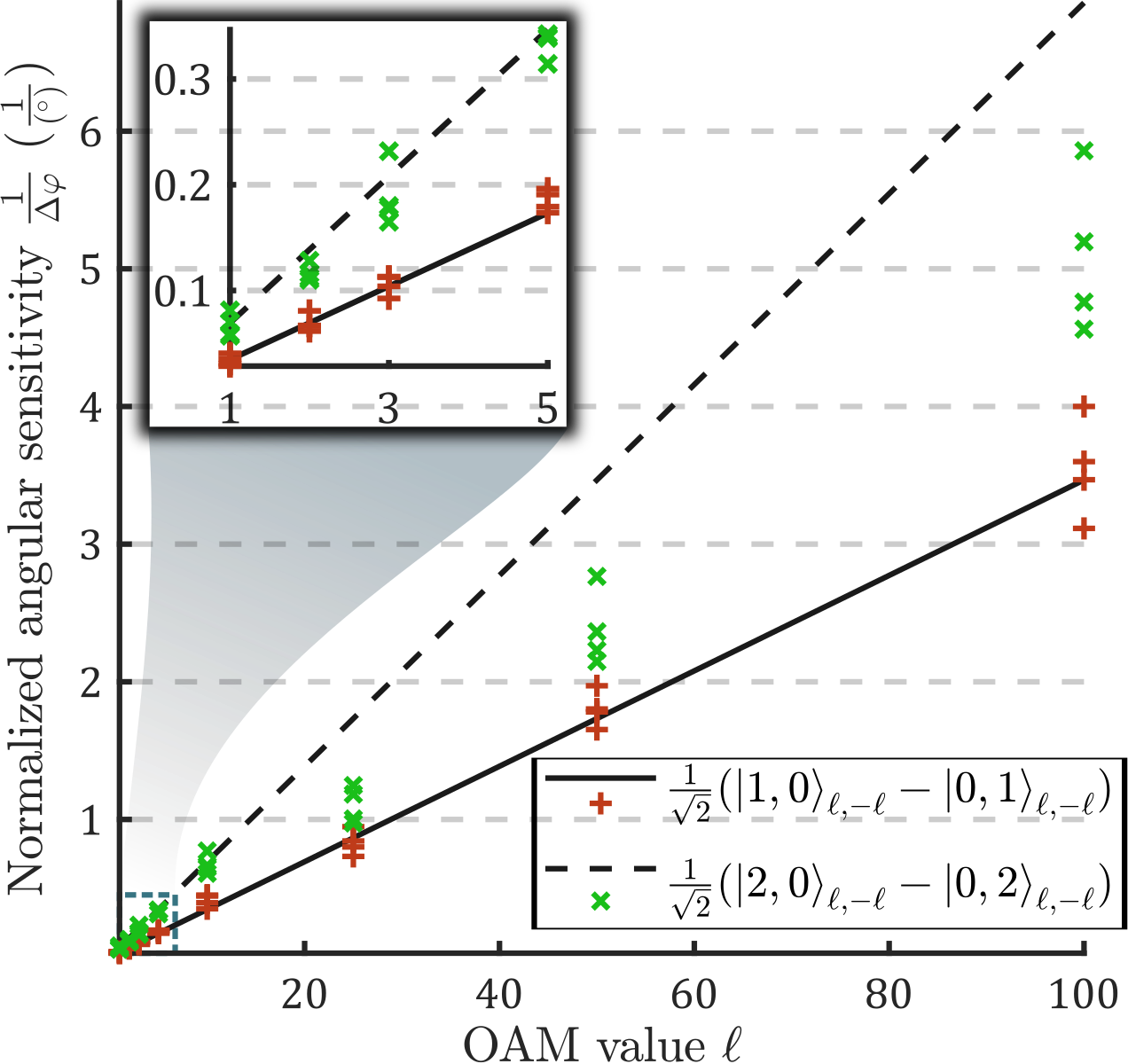}
    \caption{Measurement sensitivities of single-photon and two-photon N00N states. 
    The theoretical curves are calculated using Poissonian errors and a visibility of 0.9999 for the cosine curve introduced in eq.~\eqref{eq:fit}.
    The crosses represent the four normalized sensitivities calculated from the uncertainty values chosen from each measurement.
    The mostly linear dependence of $\nicefrac{1}{\Delta \varphi}$ on OAM follows the scaling of angular uncertainty introduced in eq.~\eqref{eq:precision_scaling}.
    }
    \label{fig:Sensitivities}
\end{figure}

Finally, we compare the sensitivities that are achievable with two-photon N00N states to single photon sensitivities, using different values of OAM.
Fig.~\ref{fig:Fisher}\,(c) and (d) show that the best angular precision tends to be found at the same values of $\varphi$ where the Fisher information is maximized.
Therefore, to quantify the achievable sensitivities with different values of $N$ and $\ell$, we take the four smallest values of angular uncertainty $\Delta \varphi$ from each measurement, close to the point of maximum Fisher information.
We then calculate the reciprocal for each of these values and define it as the sensitivity.
To be able to compare sensitivities between different measurements, we normalize them by dividing each value $\Delta \varphi$ by $\frac{\sqrt{A+D}}{A}$, which removes the dependence of $\Delta \varphi$ on the varying number of detections in each measurement. 
These values are plotted in Fig.~\ref{fig:Sensitivities}, along with theoretically expected maximum sensitivities for the used measurement scheme. 

Fig.~\ref{fig:Sensitivities} shows a good correspondence between the theoretical and measured  sensitivities.
We see the largest deviations in two-photon states with high OAM values.
This deviation is mostly due to the non-perfect visibilities of the measured interference curves, in addition to the increasing complexity of the structures and their decreasing efficiencies, causing the alignment to be more sensitive while requiring longer measuring times.
As a result, a slow misalignment over time has a larger effect on the variability of detection rates over the repeated measurements.

In the presented experiment, we created twisted one- and two-photon N00N states, and verified the scaling they enable for angular resolution and sensitivity, when increasing the photon number or OAM.
In order to verify these properties, we rotated the measuring hologram on an SLM to simulate a rotation of the light field.
Hence, the method could be directly applied to precisely aligning two rotational reference frames, e.g. in a communication channel \cite{d2013photonic}.
However, in order to apply the method for measuring rotations caused by a separate system, the probed sample needs to be the one providing an OAM-dependent phase onto our two-photon state.
This could be achieved by embedding an image rotator (e.g. a Dove prism) into the object whose rotation we want to measure, or by probing samples that interact with the N00N state by inducing an OAM-dependent phase which is contingent upon some property of the sample. 
Hence, the scheme is not restricted to only measuring rotations of a light field or reference frame, but can be used to measure any OAM dependent phase changes.
Additionally, since adding photons into the N00N state can be done irrespective of the aperture of the system, the increased angular resolution provided by a N00N state might be beneficial in tasks with a limited aperture size. 
However, in order to push the limits of achievable sensitivity with this measurement scheme, the system losses need to be reduced and a more appropriate estimator for the rotation angle should be devised \cite{slussarenko2017unconditional,lyons2018attosecond}.

In summary, we showed that by structuring and overlaying two photons, a high-fidelity two-photon N00N state can be created between any two high-OAM spatial structures.
With this method, we are able to bunch two photons into modes with up to OAM $100\,\hslash$, with minimal experimental complexity.
For future implementations, improving the methods efficiency would be key in pushing the achievable sensitivity.
The current losses are mostly caused by the probabilistic overlapping and separation of the photon pair, as well as the methods used for generation and detection.
However, the efficiency of the system could be increased by using methods that are, in principle, lossless for preparing and measuring the spatial modes \cite{hiekkamaki2019near}, and for combining the two photons into the same beam path \cite{labroille2014efficient,fontaine2019laguerre}.
Thus, the presented method opens a viable path for demonstrating an unconditional quantum advantage for rotation estimation, and provides a powerful tool for investigating the properties of N00N states with more complex spatial structures.

The authors thank Shashi Prabhakar, Rafael F. Barros, and Lea Kopf for fruitful discussions.
MH and RF acknowledge the support of the Academy of Finland through the Competitive Funding to Strengthen University Research Profiles (decision 301820), (Grant No. 308596), and the Photonics Research and Innovation Flagship (PREIN - decision 320165). 
MH also acknowledges support from the Magnus Ehrnrooth foundation through its graduate student scholarship.
FB acknowledges support from the National Research Council’s High Throughput Secure Networks challenge program and the Joint Centre for Extreme Photonics.
RF also acknowledges support from the Academy of Finland through the Academy Research Fellowship (Decision 332399).

\bibliography{Biblio.bib}
\end{document}

\title{Supplementary material to: \\ Photonic angular super-resolution using twisted N00N states}

\author{Markus Hiekkam\"aki}
\email{markus.hiekkamaki@tuni.fi}
\affiliation{Tampere University, Photonics Laboratory, Physics Unit, Tampere, FI-33720, Finland}

\author{Fr\'ed\'eric Bouchard}
\affiliation{National Research Council of Canada, 100 Sussex Drive, Ottawa, Ontario K1A 0R6, Canada}
\author{Robert Fickler}
\email{robert.fickler@tuni.fi}
\affiliation{Tampere University, Photonics Laboratory, Physics Unit, Tampere, FI-33720, Finland}

\maketitle
\onecolumngrid

\section{Theoretically expected results}
Derivation of detection rate and measurement uncertainty, based on the derivations in Refs. \cite{jha2011supersensitive, scully1993quantum}.
The $M$ independent $N$-photon N00N states, after rotation, can be expressed as
\begin{equation}
    \label{eq:initial_state}
    \ket{\Psi} = \prod_{i=1}^M \ket{\psi_{\varphi}}_i= \prod_{i=1}^M\frac{1}{\sqrt{2}}\left(\ket{N,0}_{\ell,-\ell;i} - e^{-2\text{i}N\ell\varphi}\ket{0,N}_{\ell,-\ell;i}\right).
\end{equation}
In a measurement, each N-photon state should be projected onto the state
\begin{equation}
\label{eq:proj_state}
    \ket{\psi_0} = \frac{1}{\sqrt{2}}\left(\ket{N,0}_{\ell,-\ell} + \ket{0,N}_{\ell,-\ell}\right).
\end{equation}
Hence, the measurement operator, identifying the number of N-photons detected out of our M independent states, is of the form
\begin{equation}
    \hat{M} = \sum^M_{i = 1} \hat{m}_i = \sum^M_{i = 1} \ket{\psi_0}_i\bra{\psi_0}_i = \sum^M_{i = 1} \frac{1}{2} \left[\ket{N,0}_i\bra{N,0}_i + \ket{N,0}_i\bra{0,N}_i + \ket{0,N}_i\bra{N,0}_i + \ket{0,N}_i\bra{0,N}_i \right],
\end{equation}
where we have defined $\ket{N,0} = \ket{N,0}_{\ell,-\ell}$.
The expectation value for the number of N-photon detections is of the form
\begin{equation}
\label{eq:complete_expc}    
\begin{aligned}
    \bra{\Psi}\hat{M} \ket{\Psi} &= \prod_{j=1}^M\bra{\psi_{\varphi}}_j \sum^M_{i = 1} \frac{1}{2} \left[\ket{N,0}_i\bra{N,0}_i + \ket{N,0}_i\bra{0,N}_i + \ket{0,N}_i\bra{N,0}_i + \ket{0,N}_i\bra{0,N}_i \right] \prod_{p=1}^M\ket{\psi_{\varphi}}_p.
\end{aligned}
\end{equation}
In eq.~\eqref{eq:complete_expc}, only the $j = i = p$ part of each term can differ from unity.
Leaving us with
\begin{equation}
\label{eq:expec}
\begin{aligned}
    \bra{\Psi}\hat{M} \ket{\Psi}  = & \frac{1}{4} \sum^M_{i = 1} \left[ \bra{N,0}_i - e^{2\text{i}N\ell\varphi}\bra{0,N}_i\right] [\ket{N,0}_i\bra{N,0}_i + \ket{N,0}_i\bra{0,N}_i + \ket{0,N}_i\bra{N,0}_i \\ & 
    + \ket{0,N}_i\bra{0,N}_i ] \left[ \ket{N,0}_i - e^{-2\text{i}N\ell\varphi}\ket{0,N}_i\right] \\
     = & \frac{1}{4} \sum^M_{i = 1} \left[ 1 - e^{-2\text{i}N\ell\varphi} - e^{2\text{i}N\ell\varphi} + 1 \right] \\
    = &  \frac{M}{2} \left( 1 - \cos(2N\ell\varphi) \right). 
\end{aligned}
\end{equation}

\begin{figure}[t]
    \centering
    \includegraphics[width = 0.6\textwidth]{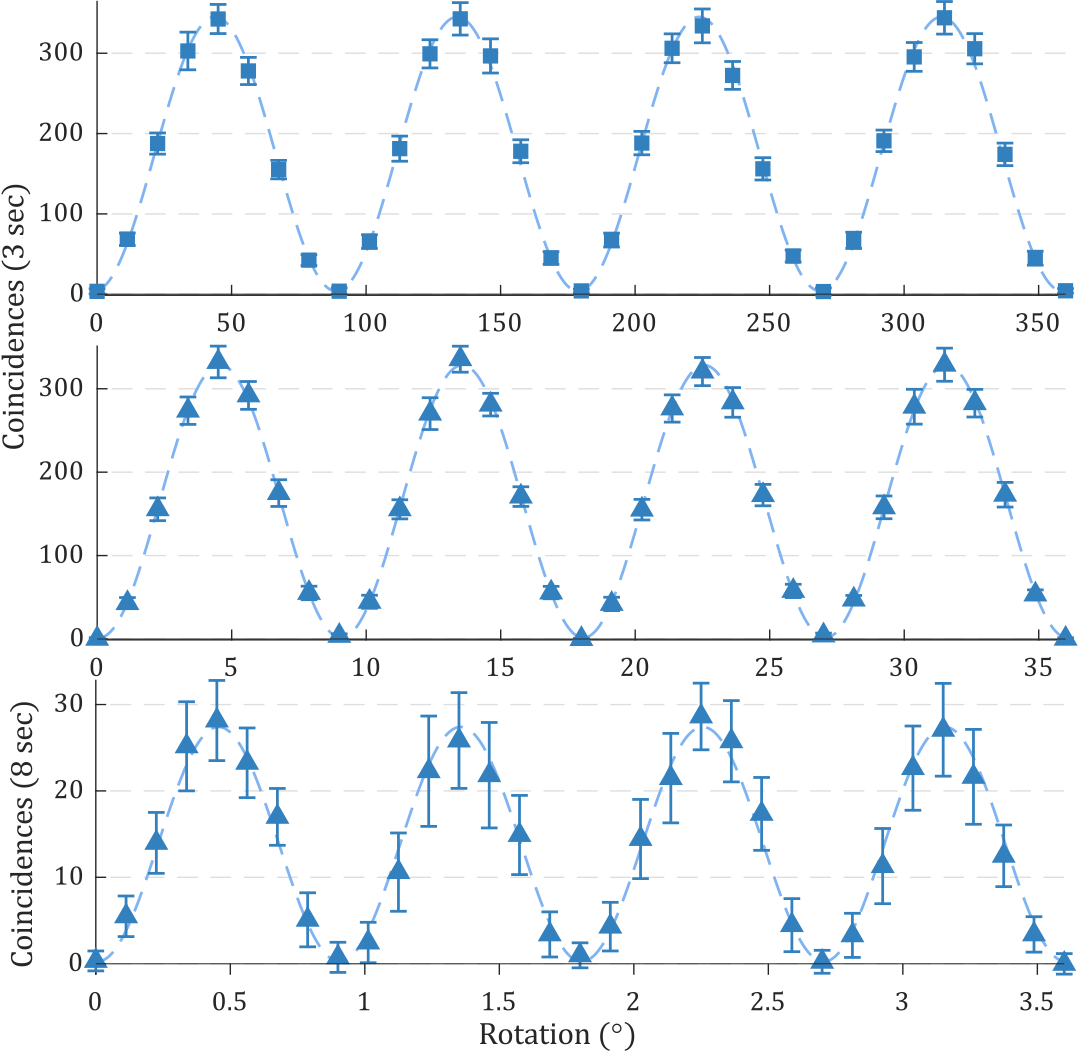}
    \caption{Two-photon N00N state measurements for $|\ell|$ = 1, $|\ell|$ = 10, and $|\ell|$ = 100. 
        In contrast to the measurements presented in the main text (Fig.~2), here the photons were projected onto the same superposition structure.
        The error bars were calculated from around 25 repetitions for the $|\ell|$ = 1 and 10 measurements, and from 19 repetitions for the $|\ell|$ = 100 measurements.
        Accidentals have been subtracted.
        In comparison to the data shown in the main article, the positions of the peaks and troughs are flipped.
        The visibilites for each curve (from weighted nonlinear least squares fits) are $0.978\pm0.002$ for $|\ell|$ = 1, $0.997\pm0.001$ for $|\ell|$ = 10, and $0.982\pm0.009$ for $|\ell|$ = 100.
        Where the errors are standard errors, calculated from the confidence intervals of the fitting parameters.}
    \label{fig:20_plot}
\end{figure}

To then calculate the theoretical detection uncertainty 
\begin{equation}
    \left|\Delta \varphi \right| = \frac{\expec{\Delta {\hat{M}}}}{\left| \nicefrac{\partial \expec{\hat{M}}}{\partial \varphi}\right|},
\end{equation}
we first need to calculate 
\begin{equation}
\begin{aligned}
\expec{\Delta {\hat{M}}}^2 & = \expec{\hat{M}^2} - \expec{\hat{M}}^2 
                                = \bra{\Psi} \left(\sum^M_{i = 1} \hat{m}_i\right) \left(\sum^M_{j = 1} \hat{m}_j \right) \ket{\Psi} -\frac{M^2}{4} \left( 1 - \cos(2N\ell\varphi) \right)^2 \\
                                & = \bra{\Psi} \left(\sum^M_{i = 1} \hat{m}_i\right) \left(\sum^M_{\substack{j = 1\\j \neq i}} \hat{m}_j \right) \ket{\Psi} + \bra{\Psi} \left(\sum^M_{k = 1} \hat{m}_k \hat{m}_k \right) \ket{\Psi}-\frac{M^2}{4} \left( 1 - \cos(2N\ell\varphi) \right)^2.
\end{aligned}
\end{equation}
Since $\hat{m}_i\hat{m}_i = \hat{m}_i$, and the same conditions for the different terms apply as in eq.~\eqref{eq:complete_expc}, the above equation simplifies to
\begin{equation}
\begin{aligned}
    \expec{\Delta {\hat{M}}}^2 & = \sum^M_{i = 1} \bra{\psi}_i \hat{m}_i\ket{\psi}_i \sum^M_{\substack{j = 1\\j \neq i}} \bra{\psi}_j \hat{m}_j\ket{\psi}_j +   \sum^M_{k = 1} \bra{\psi}_k \hat{m}_k \ket{\psi}_k - \frac{M^2}{4} \left( 1 - \cos(2N\ell\varphi) \right)^2 \\
    & = \left(\sum^M_{i = 1} \frac{1}{2}\left( 1 - \cos(2N\ell\varphi) \right)\right)\left(\sum^M_{\substack{j = 1\\j \neq i}} \frac{1}{2}\left( 1 - \cos(2N\ell\varphi) \right)\right) + \sum^M_{k = 1} \frac{1}{2}\left( 1 - \cos(2N\ell\varphi) \right) - \frac{M^2}{4} \left( 1 - \cos(2N\ell\varphi) \right)^2 \\
    & = -\frac{M}{4}\left( 1 - \cos(2N\ell\varphi) \right)^2 + \frac{M}{2}\left( 1 - \cos(2N\ell\varphi) \right) \\ 
    & = M\sin^2(N\ell \varphi)\left( 1 - \sin^2(N\ell \varphi)\right).
\end{aligned}
\end{equation}
Next, the derivative of the expectation value is
\begin{equation}
\frac{\partial \expec{\hat{M}}}{\partial \varphi} = \frac{\partial}{\partial \varphi} \frac{M}{2} \left( 1 - \cos \left( 2\ell N \varphi \right) \right) = MN\ell \sin\left(2 \ell N \varphi \right), 
\end{equation}
giving us the relation
\begin{equation}
\left|\Delta \varphi \right| = \frac{\expec{\Delta {\hat{M}}}}{\left| \nicefrac{\partial \expec{\hat{M}}}{\partial \varphi}\right|} = 
\frac{\sqrt{M}\left|\sin \left(\ell N \varphi \right) \sqrt{\left(1 -\sin^2 \left(\ell N \varphi \right) \right)}\right|}{2MN\ell \left|\sin\left(\ell N \varphi \right) \cos\left(\ell N \varphi \right)\right|} = \frac{1}{2\sqrt{M}N\ell}.
\end{equation}

\section{Measurement in experiment}
In our measurements we project the two photons onto two orthogonal states while rotating the measurement holograms.
Meaning that we combined the beam rotation and state projection into both measurement holograms and effectively project the photons onto the superposition states $\hat{a}^{\dagger}_{\varphi, \pm} = \frac{1}{\sqrt{2}}(\hat{a}^{\dagger}_{\ell} \pm e^{-2\text{i}N\ell\varphi} \hat{a}^{\dagger}_{-\ell})$, separately.

It is worth noting that projecting the two photons onto the same rotating structure $\hat{a}^{\dagger}_{\varphi} = \frac{1}{\sqrt{2}}(\hat{a}^{\dagger}_{\ell} + e^{-2\text{i}N\ell\varphi}\hat{a}^{\dagger}_{-\ell})$ is equally valid.
In our measurement scheme, in which we probabilistically split the two photons, the main differences between the results obtained from these projections is a flipping of the positions of the peaks and troughs in the measured data, and the fact that the projection on identical structures would produce a similar fringe pattern even without bunching (only the height of the peaks increases with bunching).
This switching between the peak locations can be seen by writing the two-photon version of the state $\ket{\psi_\varphi}$, given in eq.~\eqref{eq:initial_state}, in the MUB spanned by the petal structures $M_1$ and $M_2$
\begin{equation}
    \ket{\psi_\varphi} = \frac{1}{2^{\frac{3}{2}}} \left(\ket{2,0}_{M_1,M_2} + \ket{0,2}_{M_1,M_2}\right)(1-e^{-2\text{i}N\ell\varphi}) - \frac{1}{2}\ket{1,1}_{M_1,M_2}(1+e^{-2\text{i}N\ell\varphi}).
\end{equation}
To demonstrate this, Fig.~\ref{fig:20_plot} contains rotation measurements for $|\ell| = \{1,10,100\}$ twisted N00N states projected onto the same petal structure.

\section{Fisher information}

\begin{figure}[t]
    \centering
    \includegraphics[width = \textwidth]{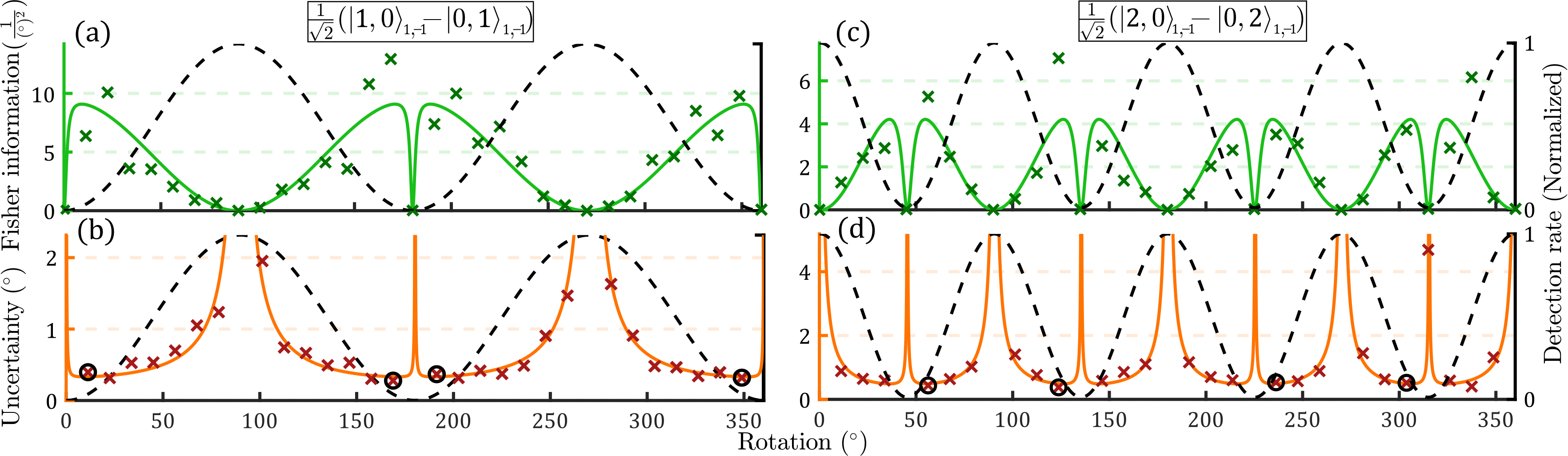}
    \caption{Fisher information and angular uncertainty of OAM N00N states for $\ell = 1$. 
    On the upper row, the continuous green line is the Fisher information $F$ multiplied by the estimate for the total number of heralded single photons (or two-photons) before losses ($M_T$).
    The green crosses are the reciprocal of the measured variance which should follow the Cram\'er-Rao bound $\text{Var}(\varphi) \geq \nicefrac{1}{M_TF}$ \cite{polino2020photonic}.
    On the bottom row, the continuous curves are calculated using eq.(3) from the main text and Poissonian errors calculated from the fit. 
    The red crosses are the experimentally determined precision values calculated using eq.(3) from the main text.
    (a) and (b) were calculated from the heralded single photon data for $|\ell| = 1$, and (c) and (d) contain two-photon data for $|\ell| = 1$.
    In all graphs, the black dashed lines depict the interference curves for reference. 
    On the bottom row, the precision values that have been circled are used for calculating the respective sensitivities in Fig.~4 of the main text.}
    \label{fig:L1Fisher}
\end{figure}

We calculate the Fisher information from the fits to our measurement data.
Since the fits are of the form $\frac{A}{2}\left(1-\cos \left( 2N\ell \frac{\pi}{180^{\circ}} \varphi-c \right)\right) + D$, the measurement probabilities are simply 
$P_1 = \frac{\eta}{A+D}\left[\frac{A}{2}\left(1-\cos \left( 2N\ell \frac{\pi}{180^{\circ}} \varphi-c \right)\right) + D\right]$ and $P_2 = 1 - \frac{\eta}{A+D}\left[\frac{A}{2}\left(1-\cos \left( 2N\ell \frac{\pi}{180^{\circ}} \varphi-c \right)\right) + D\right]$.
Here, $P_1$ corresponds to a successful detection of a heralded single photon, or a two-photon state, and $P_2$ takes into account the estimate of the losses in the system while ensuring that $P_1 + P_2 = 1$.
The efficiency of the system $\eta$ was estimated by measuring the losses with a laser.
This was done separately for different spatial structures.
Then, this estimate is multiplied by the efficiencies of our single photon avalanche diodes (74~\% and 75~\%, Laser Components COUNT T) at 810~nm.
This of course excludes the efficiency of our two-photon source and the slight difference in alignment when switching from the laser to the single photon source.
However, as can be seen from Fig.~3 in the main article, the analysis is sufficient for a reasonable estimate of the Fisher information at different values of $\varphi$.
The measured efficiencies for $\ell = 1$ are 2.6~\% for the channel used for heralded single photons and 2.01~\% on average between the four possible input and output combinations.
These losses stem from the modulation efficiency of our spatial light modulators (SLMs), which was around 70-75~\%, the beamsplitters that had a splitting ratio close to 50:50, the coupling efficiency at the last SMF, and losses caused by the mode carving and intensity flattening methods \cite{bolduc2013exact, bouchard2018measuring}.
The varying efficiencies between different combinations of input and output fibers were caused by imperfections in our imaging systems, imperfect splitting ratios of the beamsplitters, and slight differences in alignment. 

\begin{figure*}[t]
    \centering
    \includegraphics[width = \textwidth]{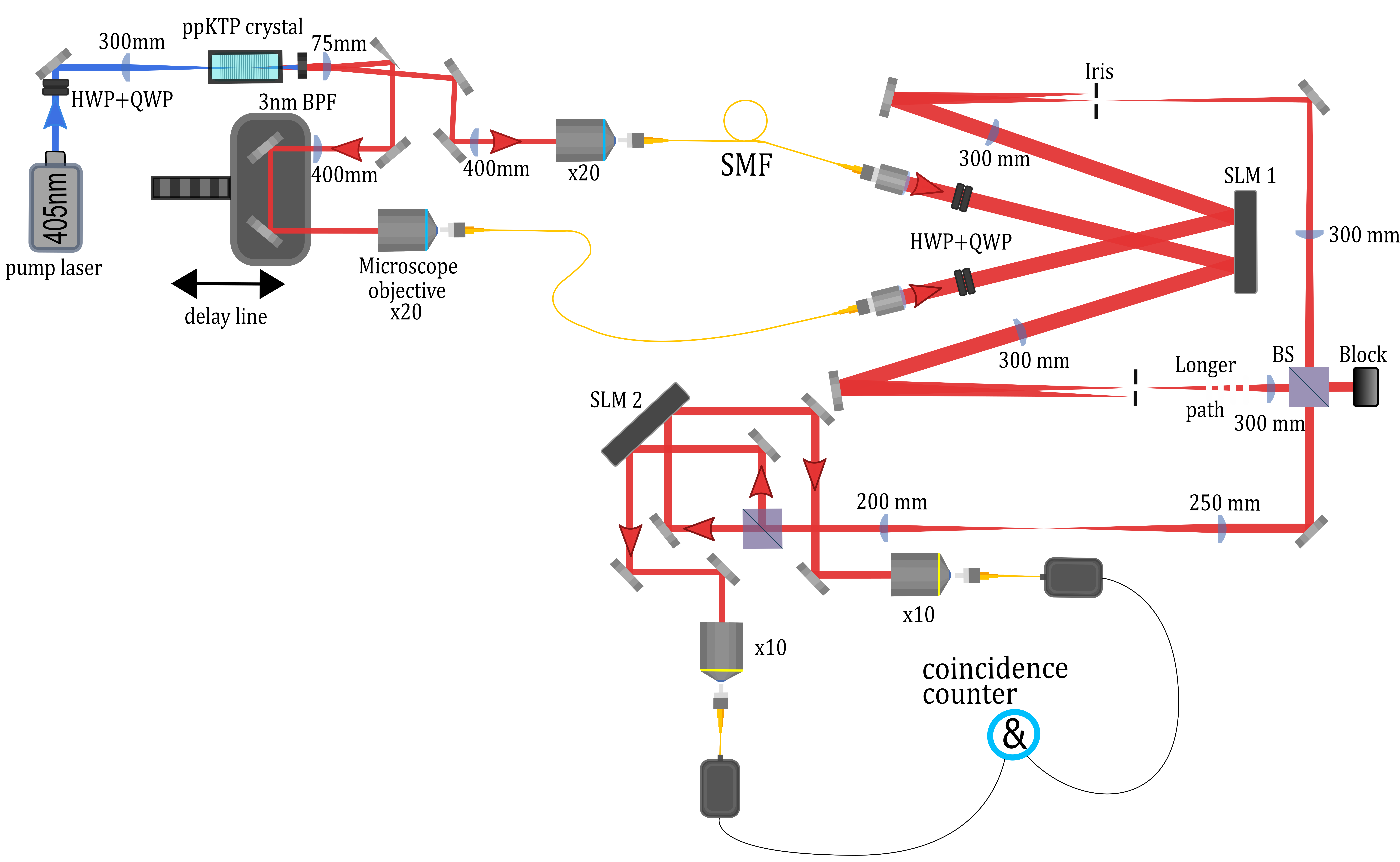}
    \caption{Detailed drawing of the used setup. The photon pair source can be found in the top left corner. The drawing lists the focal lengths of the used plano-convex lenses in addition to the magnifications of the used microscope objectives. SMF stands for single mode fiber, SLM stands for spatial light modulator, HWP stands for half-wave plate, and QWP for a quarter-wave plate. The ``longer path'' means that a physically longer path was used at that position (achieved with a few mirrors). This is because the path lengths for each photon from out-coupling to the beamsplitter should be within the coherence length to achieve bunching \cite{HOM_orig}. 
    }
    \label{fig:setup}
\end{figure*}

We calculate the total efficiencies as $\eta_{1,1} = 0.026\times 0.75\times 0.74$ and $\eta_{1,2} = 0.0201^2\times 0.74\times 0.75$ for the single photon and two-photon measurements with $\ell = 1$, respectively.
More specifically, for every component the losses of the system have a two-fold effect in the two-photon case (e.g. $0.5^2  = 0.25$ for the first beamsplitter), hence we estimate the efficiency of the system by squaring the average efficiency over all channels.
We performed the same calculations for $\ell = 100$ which had an efficiency of 0.63~\% for the single photon channel and 0.29~\% on average between all channels.
With these probabilities, the Fisher information is then calculated as \cite{polino2020photonic}
\begin{equation}
\begin{aligned}
    &F(\varphi) = \sum_{i = 1} \frac{1}{P_i} \left(\frac{\partial P_i}{\partial \varphi}\right)^2 \\ 
    &= \left[ \frac{\frac{\eta}{A+D}\left(\frac{\pi AN\ell}{180^{\circ}}\sin(2N\ell \frac{\pi}{180^{\circ}} \varphi - c)\right)^2}{\frac{A}{2}(1-\cos(2N\ell \frac{\pi}{180^{\circ}} \varphi - c))+D} - \frac{\left(\frac{\eta}{A+D} \frac{\pi AN\ell}{180^{\circ}}\sin(2N\ell \frac{\pi}{180^{\circ}} \varphi - c)\right)^2}{1-\frac{\eta}{A+D}\left[\frac{A}{2}(1-\cos(2N\ell \frac{\pi}{180^{\circ}} \varphi - c))+D\right]}\right].
\end{aligned}
\end{equation}

Finally, in order to compare the calculated Fisher information directly to our measurement data, we need the number of independent repetitions of the measurement $M_{T}$ \cite{polino2020photonic}. 
Using our system losses $\eta$, and the measured maximum detection rates, we estimate $M_T$ to be $M_T = \frac{A+D}{\eta}$.
One example of the calculated Fisher information for $\ell = \pm 1$ can be seen in Fig.~\ref{fig:L1Fisher}.

\begin{figure}[htb]
    \centering
    \includegraphics[width = 0.5\textwidth]{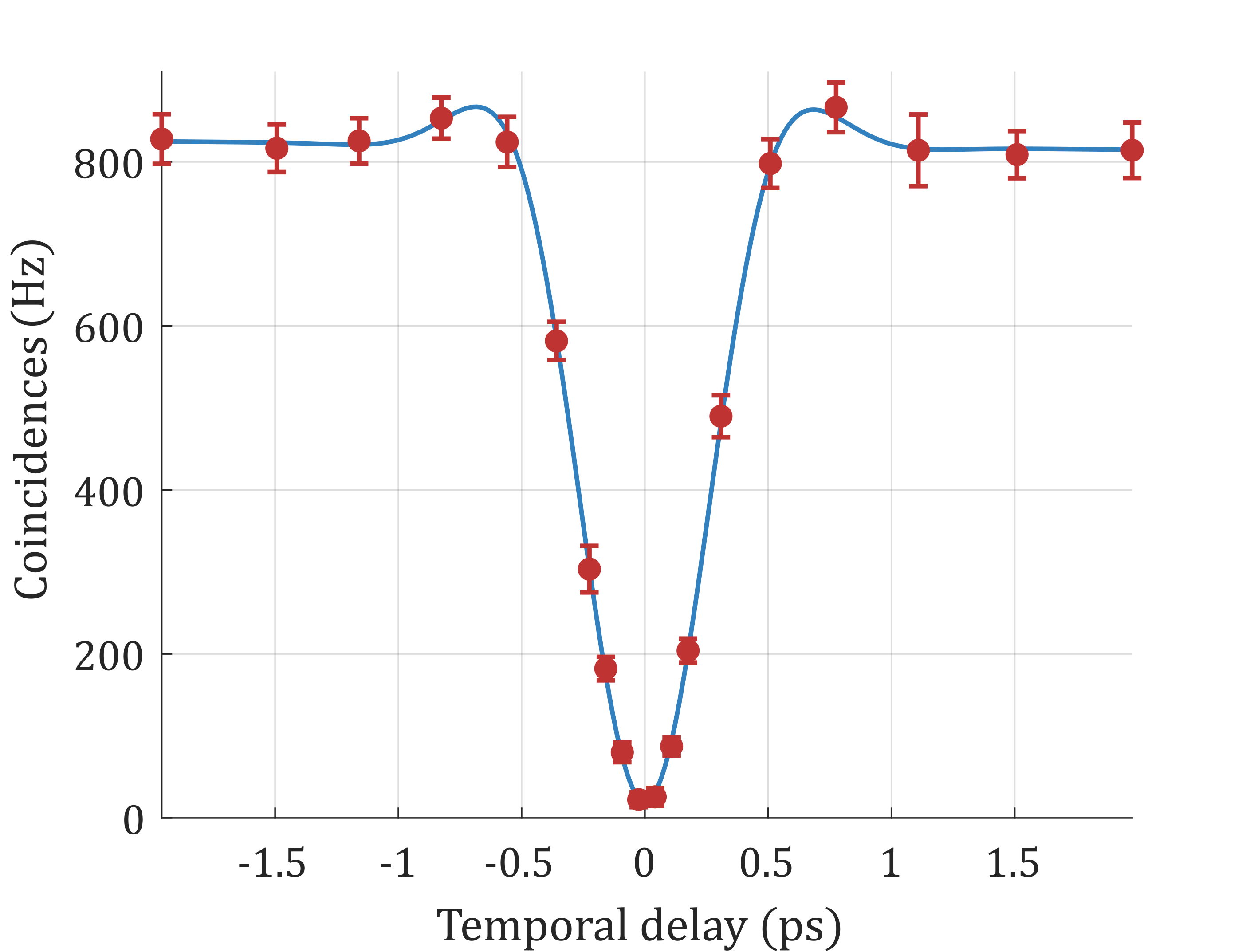}
    \caption{A scan of a Hong-Ou-Mandel dip to verify the two photon bunching into the OAM structures. Accidental coincidences have been removed from the data, and from a non-linear least squares fit to the data, a Hong-Ou-Mandel visibility of $0.976 \pm 0.002$ was calculated (where the error is a standard error calculated from the parameter confidence intervals of the fit).
The error bars were calculated from 50 repetitions of each measurement.
For more details on the calculation of the accidental rates, the fit, and the visibility, see Ref.~\cite{hiekkamakiHighDim}.}
    \label{fig:dip}
\end{figure}

\section{Experimental setup}
A detailed drawing of the experimental setup is shown in Fig.~\ref{fig:setup}.
In the source, we pump a periodically poled potassium titanyl phosphate (ppKTP, Type 0, 12~mm long) crystal with a 405~nm laser that has a linewidth below 0.06~nm  and a continuous-wave free-space power of 135~mW. 
The pump is then focused in to the crystal that down-converts the high energy photons into two 810~nm photons.
The photons are then filtered through a 3~nm bandpass filter and separated from each other using the momentum correlations of the photon pair by placing a sharp edged mirror into the Fourier plane of the crystal.
The output of the crystal is then imaged onto two microscope objectives which are used to couple each photon into a single mode fiber (SMF).
Before coupling, one of the photons goes through a computer-controlled delay line, which is used to adjust the temporal overlap of the photon pair and realized using a motorized translation stage.

The two photons are then directed onto a spatial light modulator (Holoeye Pluto-2, wavefront corrected using the method in Ref. \cite{jesacher2007wavefront}) where they are prepared independently in the wanted spatial structures.
The pair is then imaged through identical imaging systems and overlapped probabilistically using a beamsplitter.
After this, the pair is split by a second beamsplitter and imaged onto a second SLM (Holoeye Pluto-2, wavefront corrected) where two holograms, in conjuction with two SMFs, are used to independently measure the spatial structures \cite{bouchard2018measuring, mair2001entanglement}.
Finally, the photons are fed into avalanche photodiodes (Laser Components Count-T modules), from which the signal is sent into a coincidence counter (IDQ ID900) that postselects on photon pairs arriving within a 1~ns coincidence window.

\section{Bunching curve}
To further demonstrate that the photons bunch into two different OAM structures, we measured a Hong-Ou-Mandel type interference curve.
In the measurement, the two photons were prepared in an OAM = $1\,\hslash$ two-photon N00N state, using the methods specified in the main text. 
The photons were then projected independently onto the two orthogonal OAM states with $\ell = 1$ and $\ell = -1$, respectively.
Then the indistinguishability of the two photons was varied by changing their temporal overlap with the delay line in the two-photon source (see Fig.~\ref{fig:setup}).
The measured dip can be seen in Fig.~\ref{fig:dip}.
From the measured data, it can be seen that the two photons bunch into the same spatial structure at zero delay and hence the coincidence counts drop significantly.

\bibliography{Biblio.bib}